\documentclass[13 pt]{article}
\usepackage{amssymb}

\newcommand{\betah}{\hat{\beta}}

\newcommand{\beq}{\begin{equation}}
\newcommand{\eeq}{\end{equation}}
\newcommand{\bea}{\begin{eqnarray*}}
\newcommand{\eea}{\end{eqnarray*}}
\newcommand{\beqa}{\begin{eqnarray}}
\newcommand{\eeqa}{\end{eqnarray}}

\begin{document}

\newfont{\elevenmib}{cmmib10 scaled\magstep1}%

\newcommand{\Title}[1]{{\baselineskip=26pt \begin{center}
            \Large   \bf #1 \\ \ \\ \end{center}}}
\hspace*{2.13cm}%
\hspace*{1cm}%
\newcommand{\Author}{\begin{center}\large
           Pascal Baseilhac\footnote{
baseilha@phys.univ-tours.fr}
\end{center}}
\newcommand{\Address}{{\baselineskip=18pt \begin{center}
           \it Laboratoire de Math\'ematiques et Physique Th\'eorique CNRS/UMR 6083,\\
Universit\'e de Tours, Parc de Grandmont, 37200 Tours, France
      \end{center}}}
\baselineskip=13pt

\bigskip
\vspace{-1cm}

\Title{An integrable structure related with tridiagonal
algebras}\Author

\vspace{- 0.1mm}
 \Address

\vskip 0.6cm

\centerline{\bf Abstract}\vspace{0.3mm}  \vspace{1mm}
The standard generators of tridiagonal algebras, recently
introduced by Terwilliger, are shown to generate a new (in)finite
family of mutually commuting operators which extends the
Dolan-Grady construction. The involution property relies on the
tridiagonal algebraic structure associated with a deformation
parameter $q$. Representations are shown to be generated from a
class of quadratic algebras, namely the reflection equations. The
spectral problem is briefly discussed. Finally, related massive
quantum integrable models are shown to be superintegrable.

\vspace{0.1cm}  {\small PACS: 02.20.Uw; 11.30.-j; 11.25.Hf;
11.10.Kk}
\vskip 0.8cm

\vskip -0.6cm


\section{Introduction}
In the Liouville-Arnold sense, a classical Hamiltonian system with
$N$ degrees of freedom is called completely integrable if there
exists a set of $N$ functionally independent globally defined
integrals of motion in involution, i.e. mutually commuting
regarding to the Poisson bracket. It is said to be superintegrable
if it admits more than $N$ integrals of motion. By analogy, a
quantum system with $N$ degrees of freedom is completely
integrable if there exists a set of $N-1$ operators, say ${\cal
H}_n$, $n=1,...,N-1$, together with the Hamiltonian ${\cal H}$
which are mutually commuting. If there exists $p-$additional
operators ${\cal I}_m$, $m=1,...,p$ where $1\leq p\leq N$ such
that
\beqa [{\cal H}_n,{\cal H}_m]=0 \qquad \quad \mbox{and}\qquad
\quad[{\cal H},{\cal I}_m]=0 \label{invol}\eeqa
for all $n,m$, this quantum system is said to be superintegrable.
Although not necessary for superintegrability, the additional
operators ${\cal I}_m$ can be in involution too i.e. $[{\cal
I}_n,{\cal I}_m]=0$.

\vspace{1mm}

The purpose of this paper is to introduce a new (in)finite set of
mutually commuting quantities, whose involution property will be
indentified with the defining relations for the tridiagonal
algebras recently introduced and studied by Terwilliger
\cite{Ter03} (see also \cite{Ter93},\cite{Ter01}). We will also
show that the underlying structure is closely related with
$U_{q^{1/2}}(\widehat{sl_2})$, and coincides with the Dolan-Grady
one \cite{DG} for $q=1$. As a consequence, we will show that
related quantum integrable models (open spin chains, sine-Gordon
field theory,...) enjoy superintegrability.\vspace{1mm}

This paper is organized as follows. In Section 2, after recalling
some definitions we propose a new realization of tridiagonal pairs
in terms of $U_{q^{1/2}}(\widehat{sl_2})$ generators. In
particular, the defining relations of the tridiagonal algebra are
shown to be invariant under the coproduct homomorphism of
$U_{q^{1/2}}(\widehat{sl_2})$. Based on the tridiagonal algebraic
structure, we propose a finite set of conserved quantities ${\cal
I}_n$ in involution. In Section 3, an alternative construction of
the conserved quantities ${\cal I}_n$ is described. Their
generating function is expressed in terms of general solutions of
a class of quadratic algebras, namely the reflection equations. In
Section 4, , we argue that the corresponding spectral problem is
 related with a system of $N$ partial $q-$difference
equations. For $N=1$, the eigenvalue is expressed in terms of
solutions of Bethe equations and enjoys a remarkable symmetry
property. Examples of massive quantum integrable models with
(in)finite degrees of freedom are considered in Section 5, where
the quantities ${\cal I}_n$ are explicitly identified. Concluding
remarks follow in the last Section.

\section{Tridiagonal algebraic structure and involution}
Tridiagonal algebras have been introduced and studied in
\cite{Ter93,Ter01,Ter03}, where they first appeared in the context
of $P-$ and $Q-$polynomial association schemes. We will here
consider a slightly simpler form of the tridiagonal algebraic
structure, which nevertheless possesses all interesting features
for further analysis. For more informations about these algebras,
we report the reader to \cite{Ter03} where precise definitions can
be found. Let us consider the tridiagonal (associative) algebra
$\mathbb{T}$ with unity generated by two operators (called
standard generators) acting on a vector space $V$, say ${\textsf
A}:V\rightarrow V$ and ${\textsf A}^*:V\rightarrow V$, subjects to
the tridiagonal relations
\beqa \big[{\textsf A},{\textsf A}^2{\textsf A}^*+{\textsf
A}^*{\textsf A}^2-(q+q^{-1}){\textsf A}{\textsf A}^*{\textsf
A}-\rho{\textsf A}^*\big]&=&0\ ,\nonumber\\
\big[{\textsf A}^*,{{\textsf A}^*}^2{\textsf A}+{\textsf
A}{{\textsf A}^*}^2-(q+q^{-1}){\textsf A}^*{\textsf A}{\textsf
A}^*-\rho{\textsf A}\big]&=&0\ \label{Talg}\eeqa
where $q$ is a deformation parameter (assumed to be not a root of
unity) and $\rho$ is a fixed scalar. According to
[\cite{Ter03},Theorem 3.10] ${\textsf A},{\textsf A}^*$ is a
tridiagonal (TD) pair as defined in [\cite{Ter01}, Definition
1.1], which complete classification is still an open problem.
Among the known examples of TD pairs, one finds a subset such that
${\textsf A},{\textsf A}^*$ have eigenspaces of dimension one.
These are called Leonard pairs, classified in \cite{TerLP01}. In
particular, they satisfy (for details, see \cite{TerAW03}) the
so-called Askey-Wilson (AW) relations first introduced by Zhedanov
\cite{Zhed92}. Other examples of TD pairs can be found in
\cite{Ter03,TerIto04}: for $\rho=0$ in which case (\ref{Talg})
reduce to $q-$Serre relations or for $q=1$, $\rho=16$ which leads
to Dolan-Grady relations \cite{DG}.

Here, our aim is to construct a new family of TD pairs in terms of
$U_{q^{1/2}}(\widehat{sl_2})$ satisfying (\ref{Talg}). To do that
we introduce five generators $Q_\pm,{\overline Q}_\pm$ and $H$
subjects to the algebraic relations
\beqa &&q^{-1/2}Q_\pm{\overline Q}_\pm - q^{1/2}{\overline Q}_\pm
Q_\pm =0\ ,\nonumber\\
&&q^{1/2}Q_\pm{\overline Q}_\mp - q^{-1/2}{\overline Q}_\mp
Q_\pm =\frac{q^{\pm 2H}-1}{q^{1/2}-q^{-1/2}}\ ,\nonumber\\
&&q^{\epsilon H}Q_\pm = q^{\pm \epsilon}Q_\pm q^{\epsilon H}\
,\quad q^{\epsilon H}{\overline Q}_\pm = q^{\pm
\epsilon}{\overline Q}_\pm q^{\epsilon H}\ \label{Uqsl2}\eeqa
and the $q-$Serre relations
\beqa Q_\pm^3 Q_\mp -(1+q+q^{-1})Q_\pm^2Q_\mp Q_\pm +
(1+q+q^{-1})Q_\pm Q_\mp Q_\pm^2 - Q_\mp Q_\pm^3 =0\ ,\nonumber \\
{\overline Q}_\pm^3 {\overline Q}_\mp -(1+q+q^{-1}){\overline
Q}_\pm^2{\overline Q}_\mp {\overline Q}_\pm +
(1+q+q^{-1}){\overline Q}_\pm {\overline Q}_\mp {\overline
Q}_\pm^2 - {\overline Q}_\mp {\overline Q}_\pm^3 =0\ .\nonumber
\eeqa
Using the standard definitions \cite{Chari94},  it is an exercise
to show that $Q_\pm,{\overline Q}_\pm$ and $H$ actually generate
the quantum affine Kac-Moody algebra
$U_{q^{1/2}}(\widehat{sl_2})$. Based on the results of \cite{qDG},
we propose\,\footnote{This realization  is different from the one
proposed by Terwilliger and Ito in \cite{TerIto04} in which case
$\rho=0$, i.e. (\ref{Talg}) reduce to $q-$Serre relations.} the
following TD pair:
\beqa{\textsf A}=\frac{1}{c_0}Q_+ + {\overline Q}_- \qquad
\mbox{and} \qquad {\textsf A}^*=Q_- + \frac{1}{c_0}{\overline
Q}_+\ ,\label{TDinit}\eeqa
together with the scalar
\beqa \rho=\frac{(q^{1/2}+q^{-1/2})^2}{c_0}\ .\label{rho}\eeqa
It is straightforward to check using the defining relations above
(\ref{Uqsl2}) that (\ref{TDinit}) satisfy the tridiagonal
relations (\ref{Talg}). Furthermore, we have found that the
slightly more general TD pair ${\tilde{\textsf A}}={\textsf
A}+\epsilon_+q^{H}$, ${\tilde{\textsf A}}^*={\textsf
A}^*+\epsilon_-q^{-H}$ also satisfies (\ref{Talg}) for any values
of the parameters $\epsilon_\pm$. This is not surprising as we
will see later on.\vspace{1mm}

The Hopf algebraic structure of $U_{q^{1/2}}(\widehat{sl_2})$ can
now be used to construct a whole family of TD pairs. Indeed, the
coproduct $\Delta: U_{q^{1/2}}(\widehat{sl_2}) \rightarrow
U_{q^{1/2}}(\widehat{sl_2}) \times U_{q^{1/2}}(\widehat{sl_2})$
associated with (\ref{Uqsl2}) is such that
\beqa \Delta(Q_\pm)&=& Q_\pm \otimes I\!\!I + q^{\pm H} \otimes
Q_\pm \
,\nonumber\\
 \Delta({\overline Q}_\pm)&=& {\overline Q}_\pm \otimes I\!\!I +
q^{\mp H} \otimes {\overline Q}_\pm \
,\nonumber\\
\Delta(q^{H})&=& q^H \otimes q^{H}\ .\label{coproduct}\eeqa
More generally, one defines the $N-$coproduct $\Delta^{(N)}: \
U_{q^{1/2}}(\widehat{sl_2}) \longrightarrow
U_{q^{1/2}}(\widehat{sl_2}) \otimes \cdot\cdot\cdot \otimes
U_{q^{1/2}}(\widehat{sl_2})$ as
\beqa \Delta^{(N)}\equiv (id\times \cdot\cdot\cdot \times id
\times \Delta)\circ \Delta^{(N-1)}\ \eeqa
for $N\geq 3$ with $\Delta^{(2)}\equiv \Delta$,
$\Delta^{(1)}\equiv id$. The opposite $N-$coproduct
$\Delta'^{(N)}$ is similarly defined with $\Delta'\equiv \sigma
\circ\Delta$   where the permutation map $\sigma(x\otimes y
)=y\otimes x$ for all $x,y\in U_{q^{1/2}}(\widehat{sl_2})$ is
used. Due to the homomorphism property of the coproduct (or its
opposite) $\Delta(xy)=\Delta(x)\Delta(y)$ the algebraic relations
(\ref{Uqsl2}) are invariant under the action of $\Delta$ (or
$\Delta'$). Consequently, the family of TD pairs (acting on tensor
product of irreducible representations) defined as
\beqa {\textsf A}^{(N)}=\Delta^{(N)}({\textsf A}) \qquad
\mbox{and} \qquad {{\textsf A}^*}^{(N)}=\Delta^{(N)}({{\textsf
A}^*})\qquad \mbox{for}\qquad N\geq 1 \label{TDN}\eeqa
with (\ref{TDinit}) satisfies the tridiagonal
relations\,\footnote{Note that an other family of TD pairs can
similarly be obtained using instead the opposite coproduct
$\Delta'$.} (\ref{Talg}) with (\ref{rho}). It is interesting to
notice that the TD pair ${\tilde {\textsf A}}$, ${\tilde {\textsf
A}^*}$ can also be generalized as in (\ref{TDN}). The
corresponding tridiagonal algebra ${\tilde {\mathbb{T}}}^{(N)}$ is
a (left) coideal subalgebra \cite{Letz02} of
$U_{q^{1/2}}(\widehat{sl_2})$. Indeed, one has $\Delta(a^{(N)})\in
U_{q^{1/2}}(\widehat{sl_2})\otimes {\tilde {\mathbb{T}}}^{(N)}$
for all $a\in \{I\!\!I, {\tilde {\textsf A}},{\tilde {\textsf
A}^*}\}$. As a result, this class of coideal subalgebras closes
under the tridiagonal algebraic relations
(\ref{Talg}).\vspace{2mm}

An interesting problem is whether the tridiagonal algebraic
relations (\ref{Talg}) might provide the condition of existence of
a finite set of mutually commuting quantities. Having in mind the
construction of Dolan and Grady \cite{DG} associated with
(\ref{Talg}) for $q=1$, whose generators can be expressed in terms
of the loop algebra of $sl_2$ \cite{Roan91,Date00}, this problem
seems rather natural and solvable. Inspired by \cite{DG},
\cite{qDG} we propose the following finite set of $N$ quantities
in involution, denoted ${\cal I}^{(N)}_{2n-1}$ with $n=1,...,N$,
generated by the TD pair ${\textsf A}^{(N)}$, ${{\textsf
A}^*}^{(N)}$ (\ref{TDN}) with (\ref{TDinit}):
\beqa {\cal I}_1^{(N)} &=& \kappa{\textsf A}^{(N)} +
\kappa^*{{\textsf
A}^*}^{(N)}\ ,\nonumber\\
{\cal I}_3^{(N)} &=& \kappa\Big(\big[\big[{\textsf
A}^{(N)},{{\textsf A}^*}^{(N)}\big]_q,{\textsf A}^{(N)}]_{q} +\rho
{{\textsf A}^*}^{(N)}\Big)+ \kappa^*\Big(\big[\big[{{\textsf
A}^*}^{(N)},{{\textsf A}}^{(N)}\big]_q,{{\textsf A}^*}^{(N)}]_{q}
 + \rho {{\textsf A}}^{(N)}\Big) ,\nonumber\\
{\cal I}_{5}^{(N)} &=& \cdot \cdot \cdot \ , \label{IN}\eeqa
where the $q-$commutator $[X,Y]_{q}=q^{1/2}XY-q^{-1/2}YX$ has been
introduced and $\kappa,\kappa^*$ are arbitrary parameters. As
shown in the first part of this Section, the TD pair ${\textsf
A}^{(N)}, {{\textsf A}^*}^{(N)}$ satisfies the tridiagonal
algebraic relations (\ref{Talg}). Then, it is an exercise to
check, for instance, that $\big[{\cal I}^{(N)}_{1},{\cal
I}^{(N)}_{3}\big]=0$ for any $N$. Note that the special case
$\kappa=\kappa^*$ was considered in \cite{qDG}. For general values
of $N$, it is not difficult by analogy with \cite{DG} to guess the
explicit form of the $N$ independent quantities for $n=3,...,N$
(higher order polynomials in the TD pair) which are mutually
commuting. However, due the $q-$commutator the proof of mutual
commutativity of the corresponding quantities becomes quickly
rather complicated. Consequently, in next Section we propose an
alternative method to derive these quantities.\vspace{1mm}

To see the truncation of the hierarchy at the order $N$ let us
focus on the smallest values of $N$. For $N=1$, using the
evaluation homomorphism $\pi_v:
U_{q^{1/2}}(\widehat{sl_2})\longrightarrow U_{q^{1/2}}(sl_2)$
($v\in \mathbb{C}$ is sometimes called the spectral parameter)
\beqa \pi_v[Q_\pm]= vS_\pm q^{\pm s_3/2}\ \quad \mbox{and}\qquad \
\pi_v[{\overline Q}_\pm]= v^{-1}S_\pm q^{\mp s_3/2}\qquad
\pi_\nu[H]= s_3\ ,\label{homo}\eeqa
where $\{S_\pm,s_3\}$ are the fundamental generators of the
quantum algebra $U_{q^{1/2}}(sl_2)$ satisfying the relations $
[s_3,S_\pm]=\pm S_\pm$ and $
[S_+,S_-]=(q^{s_3}-q^{-s_3})/(q^{1/2}-q^{-1/2})$, one finds that
${\textsf A}^{(1)}$, ${{\textsf A}^*}^{(1)}$ satisfy the AW
relations\,\footnote{Here we use the notations of \cite{TerAW03}
for $\rho=\rho^*$, $\gamma=\gamma^*=0$, $\eta=\eta^*=0$.}
\beqa {{\textsf A}^{(1)}}^2  {{\textsf A}^*}^{(1)} +{{\textsf
A}^*}^{(1)}{{\textsf A}^{(1)}}^2-(q+q^{-1}){{\textsf
A}^{(1)}}{{\textsf A}^*}^{(1)}{{\textsf A}^{(1)}}&=&\rho{{\textsf A}^*}^{(1)}+\omega{{\textsf A}}^{(1)}\ ,\nonumber\\
{{{\textsf A}^*}^{(1)}}^2  {{\textsf A}}^{(1)} +{{\textsf
A}}^{(1)}{{{\textsf A}^*}^{(1)}}^2-(q+q^{-1}){{{\textsf
A}^*}^{(1)}}{{\textsf A}}^{(1)}{{{\textsf
A}^*}^{(1)}}&=&\rho{{\textsf A}}^{(1)}+\omega{{\textsf
A}^*}^{(1)}\  \label{AW} \eeqa
with (\ref{rho}) and
\beqa \omega=-(v^2q^{-1/2}+v^{-2}q^{1/2})w^{(j)}/c_0\ .\eeqa
Here, $w^{(j)}= (q^{j+1/2}+q^{-j-1/2})$ denotes the eigenvalue of
the Casimir operator of  $U_{q^{1/2}}(sl_2)$ in the spin$-j$
representation. In other words, the TD pair ${\textsf A}^{(1)}$,
${{\textsf A}^*}^{(1)}$ is a Leonard pair (see \cite{TerAW03} for
details). Consequently, one can not construct an independent
quantity trilinear in ${\textsf A}^{(1)}$, ${{\textsf A}^*}^{(1)}$
in the spirit of \cite{DG}, due to (\ref{AW}). Appart from the
Casimir operator of the AW algebra \cite{Zhed92} (quartic in the
Leonard pair), there are no other independent quantities commuting
with ${\cal I}_1^{(1)}$ directly built from the TD
pair.\vspace{1mm}

For $N=2$, we have checked explicitly that the TD pair ${\textsf
A}^{(2)}$, ${{\textsf A}^*}^{(2)}$ does {\it not} satisfy the AW
relations (\ref{AW}).  As before, we expect that an {\it
independent} quantity ${\cal I}_5^{(2)}$ of fifth order in
${\textsf A}^{(2)}$, ${{\textsf A}^*}^{(2)}$ does  not exist. As
will be shown in the next section, this fact relies on the
existence of algebraic relations of fifth order in the TD pair.
More generally, there exists a set of algebraic relations
generalizing the AW ones responsible of the truncation of this
integrable hierarchy. In the next Section we will propose a
general procedure to derive all ${\cal I}_{2n-1}^{(N)}$ for
$n=1,...,N$, all algebraic relations generalizing (\ref{AW}) and
the involution property of the charges ${\cal I}_{2n-1}^{(N)}$. It
is based on the generalized quantum inverse scattering approach.

\section{Generating function}
In the spirit of the quantum inverse scattering  method, we would
like to construct an object which would provide the generating
function of the above mentioned mutually commuting quantities
${\cal I}_{2n-1}^{(N)}$, $n=1,...,N$. Following the results of
\cite{qDG}, let us consider the following quadratic algebra
(called reflection equation) which was introduced by Cherednik
\cite{Cher84}:
\beqa R(u/v)\ (K(u)\otimes I\!\!I)\ R(uv)\ (I\!\!I \otimes K(v))\
= \ (I\!\!I \otimes K(v))\ R(uv)\ (K(u)\otimes I\!\!I)\ R(u/v)\ .
\label{RE} \eeqa
This equation arises, for instance, in the context of quantum
integrable systems with boundaries \cite{Skly88}. We report the
reader to the literature on the subject for more details. For our
purpose, we restrict our attention to the trigonometric $R-$matrix
$R(u)$ which solves the Yang-Baxter equation. In the
spin$-\frac{1}{2}$ representation of
$U_{q^{1/2}}(\widehat{sl_2})$, it reads
\beqa R(u)=\sum_{i,j\in \{0,3,\pm\}}\omega_{ij}(u)\
\sigma_i\otimes\sigma_j\ , \label{R} \eeqa
where
\beqa \omega_{00}(u)&=&\frac{1}{2}(q^{1/2}+1)(u-q^{-1/2}u^{-1})\
,\qquad
\omega_{33}(u)= \frac{1}{2}(q^{1/2}-1)(u+q^{-1/2}u^{-1})\ ,\nonumber \\
\omega_{+-}(u)&=&\omega_{-+}(u)=q^{1/2}-q^{-1/2}\ \nonumber\eeqa
and $\sigma_j$ are Pauli matrices, $\sigma_\pm=(\sigma_1\pm
i\sigma_2)/2$. As shown in \cite{Skly88}, the solution of
(\ref{RE}) is not {\it unique}: a family of solutions of the
reflection equation (\ref{RE}) can be obtained from an ``initial''
solution, say $K^{(0)}(u)$,  using the so-called ``dressing''
procedure. Indeed, given the Lax operator $L(u)$ which satisfies
the Yang-Baxter algebra
\beqa R(u/v)\big({L}(u)\otimes
{L}(v)\big)=\big({L}(v)\otimes{L}(u)\big)R(u/v)\ \label{RLL}
 \eeqa
then one obtains a family of ``dressed'' reflection matrix for any
parameter $k\in \mathbb{C }$
\beqa K^{(N)}(u)\equiv L_{\verb"N"}(uk)\cdot\cdot\cdot
L_{\verb"1"}(uk)K^{(0)}(u)L_{\verb"1"}(uk^{-1})\cdot\cdot\cdot
L_{\verb"N"}(uk^{-1})\ \label{Kdressed}\eeqa
acting on the quantum space ${ V}^{(N)}\equiv
\bigotimes_{{\verb"j"}={\verb"1"}}^{\verb"N"} { V}_{\verb"j"}$
 which {\it also} solves (\ref{RE}). Following the
analysis of \cite{Skly88}, let us then introduce the elementary
solution of the ``dual'' reflection equation\,\footnote{This
equation is obtained from (\ref{RE}) by changing $u\rightarrow
u^{-1}$, $v\rightarrow v^{-1}$ and $K(u)$ in its transpose.} given
by
\beqa {K}_+(u) =\left(
\begin{array}{cc}
 uq^{{1/2}}\kappa + u^{-1}q^{-{1/2}}\kappa^*     & 0\\
0 &   uq^{{1/2}}\kappa^* + u^{-1}q^{-{1/2}}\kappa  \\
\end{array} \right) \ ,\label{Lax}\eeqa
where $\kappa^*\equiv \kappa^{-1}$ for any $\kappa\in \mathbb{C}$.
Together with (\ref{Kdressed}), it constitutes a basic element for
the construction of mutually commuting quantities. Indeed, for any
values of the spectral parameters $u,v$
\beqa  \big[t^{(N)}(u),t^{(N)}(v)\big]=0 \quad \qquad
\mbox{for}\quad \qquad t^{(N)}(u) = tr_{0}\{K_+(u)K^{(N)}(u)\}\
\label{tN}\eeqa
where $tr_{0}$ denotes the trace over the two-dimensional
auxiliary space. We refer the reader to \cite{Skly88} for details.
We are going to show that, given $N$ and the choice
$K^{(0)}(u)\equiv (\sigma_+/c_0 + \sigma_-)/(q^{1/2}-q^{-1/2})$,
the set of $N$ mutually commuting quantities ${\cal
I}_{2n-1}^{(N)}$, $n=1,...,N$ described in previous Section are
 generated from (\ref{tN}). In the following, we introduce a
 second spectral parameter
$v=kq^{1/4}$ and we take the fundamental solution of the
Yang-Baxter algebra ($L-$operator) in terms of $U_{q^{1/2}}(sl_2)$
generators:
\beqa {L}(u) =\left(
\begin{array}{cc}
 uq^{{1\over 4}}q^{s_3/2}- u^{-1}q^{-{1\over 4}}q^{-s_3/2}    &(q^{1/2}-q^{-1/2})S_-\\
(q^{1/2}-q^{-1/2})S_+    &  uq^{{1\over 4}}q^{-s_3/2}- u^{-1}q^{-{1\over 4}}q^{s_3/2}  \\
\end{array} \right) \ .\label{Lax} \eeqa
Let us assume
\beqa K^{(N)}(u) = \sum_{j\in\{0,3,\pm\}} \ \sigma_j \otimes
\Omega^{(N)}_j(u)\ \label{KN}\eeqa
where $\Omega^{(N)}_{j}(u)\in {\cal F}un(u;{\textsf
A}^{(N)},{{\textsf A}^*}^{(N)})$ are Laurent polynomials of degree
$-2N\leq d \leq 2N$ in the spectral parameter $u$, the TD pair
${\textsf A}^{(N)},{{\textsf A}^*}^{(N)}$ being subject to the
tridiagonal relations (\ref{Talg}). For $N=1$, using the
evaluation homomorphism (\ref{homo}) the result for $K^{(1)}(u)$
coincides with the - called ``dynamical'' - solution of the
reflection equation proposed in \cite{qDG} (see also \cite{BK}).
Explicitly, in terms of the Leonard pair ${\textsf
A}^{(1)},{{\textsf A}^*}^{(1)}$ it is given by (\ref{KN}) with
\beqa \Omega^{(1)}_0(u)+\Omega^{(1)}_3(u)&=& uq^{1/2}{{\textsf
A}}^{(1)} -u^{-1}q^{-1/2}{{\textsf A}^*}^{(1)} \ , \nonumber\\
\Omega^{(1)}_0(u)-\Omega^{(1)}_3(u)&=& uq^{1/2}{{\textsf
A}^*}^{(1)}
-u^{-1}q^{-1/2}{{\textsf A}}^{(1)}\ ,\nonumber \\
\Omega^{(1)}_+(u)&=& \quad   \frac{q^{1/2} u^2 + q^{-1/2}
u^{-2}}{c_0(q^{1/2}-q^{-1/2})}
 +\frac{[{{\textsf A}^*}^{(1)},{{\textsf A}}^{(1)}]_{q}}{q^{1/2}+q^{-1/2}} +\frac{\omega}{(q-q^{-1})}\ ,\nonumber \\
\Omega^{(1)}_-(u)&=&\frac{q^{1/2} u^2 + q^{-1/2}
u^{-2}}{{(q^{1/2}-q^{-1/2})}} +\frac{c_0[{{\textsf
A}}^{(1)},{{\textsf A}^*}^{(1)}]_{q}}{q^{1/2}+q^{-1/2}}
+\frac{c_0\omega}{(q-q^{-1})} \ .\label{solfin}\eeqa
>From this expression, we deduce immediately
$t^{(N=1)}=(qu^2-q^{-1}u^{-2}){\cal I}_1^{(1)}$ with
\beqa {\cal I}_1^{(1)} = \kappa{{\textsf A}}^{(1)} +
\kappa^*{{\textsf A}^*}^{(1)}\ ,\eeqa
where the evaluation homomorphism (\ref{homo}) has been used. In
particular, ${\cal I}_1^{(1)}$ is {\it linear} in the generators
${{\textsf A}}^{(1)},{{\textsf A}^*}^{(1)}$, as anticipated in the
previous Section. There are no higher quantities ${\cal
I}_{2n-1}^{(1)}$ for $n\geq 2$ generated in the expansion, a
phenomena which relies on the existence of the AW relations
(\ref{AW}) which follow from the reflection equation
\cite{qDG}.\vspace{1mm}

For $N=2$, the quantities ${\cal I}_{1}^{(2)},{\cal I}_{3}^{(2)}$
can also be exhibited in the expansion of $K^{(2)}(u)$. To show
that, we first use eqs. (\ref{TDinit}), (\ref{coproduct}) and
(\ref{TDN}) to derive
\beqa  \big[{{\textsf A}}^{(2)},\big[{{\textsf A}}^{(2)},{{\textsf
A}^*}^{(2)}\big]_q\big]_{q^{-1}}&=& \big[{{\textsf
A}}^{(1)},\big[{{\textsf A}}^{(1)},{{\textsf
A}^*}^{(1)}\big]_q\big]_{q^{-1}}\otimes I\!\!I   + q^{H}\otimes
\big[{{\textsf A}}^{(1)},\big[{{\textsf A}}^{(1)},{{\textsf
A}^*}^{(1)}\big]_q\big]_{q^{-1}}\ \nonumber\\
&+&\Big(\frac{(1-q^{-1})}{c_0}Q_+ + (1-q){\overline
Q}_-\Big)\otimes {{\textsf A}}^{(1)}{{\textsf A}^*}^{(1)}\nonumber
\\
 &+&\Big(\frac{(1-q)}{c_0}Q_+ + (1-q^{-1}){\overline
Q}_-\Big)\otimes {{\textsf A}^*}^{(1)}{{\textsf A}}^{(1)}\
\nonumber \\
&-&\frac{(q-q^{-1})}{c_0}\Big(qQ_+{\overline Q}_- -
q^{-1}{\overline
Q}_-Q_+\Big)q^{-H}\otimes {{\textsf A}^*}^{(1)}\nonumber \\
&+& \Big(\frac{(1-q)}{c_0}\big(Q_+Q_- + {\overline Q}_+{\overline
Q}_-\big)q^H + \frac{(1-q^{-1})}{c_0}\big(Q_-Q_+ + {\overline
Q}_-{\overline Q}_+\big)q^H\Big)\otimes {{\textsf A}}^{(1)} \ ,
\label{a2a2as2} \eeqa
and similarly for $\big[{{\textsf A}^*}^{(2)},\big[{{\textsf
A}^*}^{(2)},{{\textsf A}}^{(2)}\big]_q\big]_{q^{-1}}$, exchanging
$q\rightarrow q^{-1}$, ${{\textsf A}}^{(1)}\leftrightarrow
{{\textsf A}^*}^{(1)}$, $Q_+\leftrightarrow {\overline Q}_+$ and
${\overline Q}_-\leftrightarrow {Q}_-$ in the above expression.
Using the evaluation homomorphism (\ref{homo}), we replace the AW
relations (\ref{AW}) in the two first terms of (\ref{a2a2as2}). A
straightforward calculation finally shows that the diagonal
entries of $K^{(2)}(u)$ can be written in terms of
(\ref{a2a2as2}). We obtain for instance
\beqa \Omega^{(2)}_0(u)+\Omega^{(2)}_3(u)&\equiv& \
\big(qu^3+uq^{1/2}f(v)+u^{-1}\big){{\textsf A}}^{(2)}
-\big(q^{-1}u^{-3}+u^{-1}q^{-1/2}f(v)+u\big){{\textsf A}^*}^{(2)}
\nonumber \\
&&+ uq^{1/2}\frac{c_0}{q^{1/2}+q^{-1/2}}\Big(-\big[{{\textsf
A}}^{(2)},\big[{{\textsf A}}^{(2)},{{\textsf
A}^*}^{(2)}\big]_q\big]_{q^{-1}}\ +\
\frac{(q^{1/2}+q^{-1/2})^2}{c_0}{{\textsf
A}^*}^{(2)}\Big)\nonumber\\
&&- u^{-1}q^{-1/2}\frac{c_0}{q^{1/2}+q^{-1/2}}\Big(-\big[{{\textsf
A}^*}^{(2)},\big[{{\textsf A}^*}^{(2)},{{\textsf
A}}^{(2)}\big]_q\big]_{q^{-1}}\ +\
\frac{(q^{1/2}+q^{-1/2})^2}{c_0}{{\textsf A}}^{(2)}\Big) \
,\nonumber \eeqa
where
\beqa
f(v)=-\frac{2(v^2q^{-1/2}+v^{-2}q^{1/2})w^{(j)}}{q^{1/2}+q^{-1/2}}\
.\label{fv} \eeqa
The expression for $\Omega^{(2)}_0(u)-\Omega^{(2)}_3(u)$ is
derived similarly, and corresponds to the substitution ${{\textsf
A}}^{(2)}\leftrightarrow {{\textsf A}^*}^{(2)}$ in the above
equation. The mutually commuting quantities can now be read off
from the coefficients of the expansion in the spectral parameter
$u$ of (\ref{tN}). Taking the trace over the auxiliary space of
(\ref{KN}) for $N=2$ together with the expression for
$\Omega_0^{(2)}$, we find
\beqa t^{(2)}(u)=\big(q^{3/2}u^4-q^{-3/2}u^{-4}+
C(u^2,u^{-2};v)\big){\cal I}_1^{(2)} +
\frac{c_0(qu^2-q^{-1}u^{-2})}{q^{1/2}+q^{-1/2}}{\cal I}_3^{(2)} \
,\eeqa
where ${\cal I}_1^{(2)},{\cal I}_3^{(2)}$ coincide exactly with
(\ref{IN}) for $N=2$ and the Laurent polynomial $C(u,u^{-1};v)$
directly follows from (\ref{fv}). The property (\ref{tN}) implies
that the two quantities ${\cal I}_1^{(2)},{\cal I}_3^{(2)}$ are
mutually commuting. This provides an alternative derivation to the
one proposed in previous Section which is solely based on  the
fact that ${{\textsf A}}^{(2)}$, ${{\textsf A}^*}^{(2)}$ satisfy
the tridiagonal algebraic relations (\ref{Talg}).\vspace{1mm}

This technical procedure can be clearly applied for general values
of $N$. In this case, it is important to notice that the leading
terms in the asymptotic $u\rightarrow \infty$ yields to
\beqa u^{2-2N}q^{(1-N)/2}K^{(N)}(u)|_{u\rightarrow \infty}= \left(
\begin{array}{cc}
 q^{1/2}u{\textsf A}^{(N)} &  \frac{q^{1/2}u^2}{c_0(q^{1/2}-q^{-1/2})}\Delta^{(N)}(I\!\!I)\\
 \frac{q^{1/2}u^2}{q^{1/2}-q^{-1/2}}{\Delta}^{(N)}(I\!\!I) & q^{1/2}u{{\textsf A}^*}^{(N)}  \\
\end{array} \right) +
{\cal O}(u^{-2})\ .\label{asyU}\eeqa
Similarly, the asymptotic $u\rightarrow -\infty$ gives an
analogous expression exchanging ${\textsf A}^{(N)}\leftrightarrow
{{\textsf A}^*}^{(N)}$. We immediately recognize the TD pair
defined in (\ref{TDN}) which obeys (\ref{Talg}) as shown in
previous Section. Then, the mutually commuting quantities
(\ref{IN}) are obtained from the expansion of $t^{(N)}(u)$ in the
spectral parameter $u$ which can be formally written
\beqa t^{(N)}(u) =\sum_{n=1}^{N}
C^{(N)}_{2n-1}(u^{2n};u^{-2n},...;u^2,u^{-2};v){\cal
I}_{2n-1}^{(N)}\ , \label{expandt}\eeqa
where $C_{2n-1}^{(N)}$ are Laurent polynomials in $u,v$. According
to (\ref{tN}), they are in involution i.e.  $\big[{\cal
I}_{2n-1}^{(N)},{\cal I}_{2m-1}^{(N)} \big]=0$ for any $n,m\geq
1$. Such property is actually encoded in the tridiagonal relations
(\ref{Talg}). It should be stressed that the above approach does
not depend on the choice of a representation for (\ref{homo}).
Consequently, finite, infinite dimensional or cyclic
representations of TD pairs and related quantities like ${\cal
I}_{2n-1}^{(N)}$ can be easily obtained.

\section{Related spectral problem}
In the context of quantum integrable systems (see next Section),
an important problem is the diagonalization of (\ref{expandt})
which  leads to study the spectral problem associated with
(\ref{IN}) that we shall briefly discuss now. We write it as
\beqa {\cal I}^{(N)}_{2n-1}
\Psi_n^{(N)}(z_1,...,z_N)=\Lambda^{(N)}_{2n-1}\Psi_n^{(N)}(z_1,...,z_N)\
,\qquad \mbox{for}\qquad n=1,...,N\ ,\label{spectralN}\eeqa
where $\Psi_n^{(N)}(z_1,...,z_N)$ denote the eigenfunctions. For
$q$ not a root of unity and general values $N$, we expect that
this problem can be ``algebraized''\footnote{We use the definition
of \cite{Wieg95}.} i.e. part of the spectrum corresponds to
multivariable polynomial eigenfunctions. Here we do not attempt to
study in details the corresponding system of $N$ partial
$q-$difference equations, which goes beyond the scope of this
paper. Instead, let us focus on the case $N=1$ which can be solved
easily following \cite{Wieg95,Ter03} (see also \cite{qDG}).
Expressed in the weight basis, the generators of
$U_{q^{1/2}}(sl_2)$ leave invariant the linear space (of $2j+1$
dimension) of polynomials $F(z)$ of degree $2j$. The
lowest/highest weights are identified with $F_0=1$ and
$F_{2j}=z^{2j}$, respectively. One has
\beqa q^{\pm s_3/2} F(z)&=&q^{\mp j/2}F(q^{\pm 1/2}z)\ ,\nonumber\\
S_+
F(z)&=&-\frac{z}{(q^{1/2}-q^{-1/2})}\big(q^{-j}F(q^{1/2}z)-q^{j}F(q^{-1/2}z)\big)\
,\nonumber\\
S_-
F(z)&=&\frac{1}{z(q^{1/2}-q^{-1/2})}\big(F(q^{1/2}z)-F(q^{-1/2}z)\big)\
. \label{jrep}\nonumber \eeqa
In this basis of single-variable polynomials, the spectral problem
for (\ref{IN}) with $N=1$ reduces to a single second-order
$q-$difference equation of the form
\beqa a^{(1)}(z;v)\Psi(qz) + d^{(1)}(z;v)\Psi(q^{-1}z)
-v^{(1)}(z;v)\Psi(z) = \Lambda_1^{(1)}\Psi(z)\
.\label{q2diff}\eeqa
Indeed, we should keep in mind that - at this special value of $N$
- all the higher charges can be written in terms of the first one
due to the AW relations (\ref{AW}). In (\ref{q2diff}), the
coefficients are given by
\beqa a^{(1)}(z;v)&=&\frac{c_0\kappa v^{-1}q^{-j/2}z^{-1}- \kappa
vq^{-3j/2}z}{c_0(q^{1/2}-q^{-1/2})}
\ ,\nonumber \\
d^{(1)}(z;v)&=&\frac{\kappa^* v^{-1}q^{3j/2}z - c_0\kappa^*
vq^{j/2}z^{-1}}{c_0(q^{1/2}-q^{-1/2})}\ , \nonumber
\\
v^{(1)}(z;v)&=&\frac{(c_0\kappa v^{-1}q^{-j/2}-c_0\kappa^*
vq^{j/2})z^{-1}+(\kappa^* v^{-1}q^{-j/2}-\kappa vq^{j/2})z
}{c_0(q^{1/2}-q^{-1/2})}\ \label{coeffj}
 \eeqa
and
\beqa \Psi(z)=\prod_{m=1}^{M}\big(z-z_{m}\big)\ ,\label{poly}\eeqa
where $z_{m}$ denote the roots of the polynomial. Dividing
(\ref{q2diff}) by (\ref{poly}), the l.h.s. of (\ref{q2diff}) gives
a meromorphic function of $z$ and the r.h.s. becomes a constant.
Singularities of the l.h.s must be cancelled. They are located at
$z=0,\infty$ and $z=z_m$. For $z=0$, (\ref{q2diff}) vanishes
identically. For $z=\infty$ and general values of
$\kappa,\kappa^*$ on gets the constraint $M=2j$. For $z=z_m$, one
finds the following system of Bethe  equations:
\beqa \frac{a(z_l)}{d(z_l)}=\prod^{M=2j}_{m=1,m\neq l
}\frac{q^{-1}z_l-z_m}{qz_l-z_m}\qquad \mbox{for} \qquad
l=1,...,2j\ \label{Bethe}\eeqa
to which corresponds exactly $2j+1$ polynomial eigenfunctions.
Comparing the constant terms of both sides in (\ref{q2diff}) one
obtains
\beqa \Lambda_1^{(1)}=-\big(\kappa^*v^{-1}q^{-j/2+1/2}+\kappa
vq^{j/2-1/2}\big)\sum_{m=1}^{2j}z_m/c_0\ .\eeqa

Let us also mention an other way to solve the spectral problem
(\ref{spectralN}) for $N=1$, which exhibits an interesting link
with Askey-Wilson $q-$orthogonal polynomials. These Laurent
polynomials are symmetric in the variable $y$ and defined by
\beqa \textsc{P}_{n}(y)=\ _4\Phi_3 \left (
\begin{array}{c}
  q^{-n},\ abcdq^{n-1}, \ ay, \ ay^{-1}\\
  ab, \ ac, \  ad \\
\end{array}; q,q
\right)\ \label{polyAW}\eeqa
where $_4\Phi_3$ denotes the basic $q-$hypergeometric function,
$n$ is an integer and $a,b,c,d$ are arbitrary parameters. In
particular, the zeros of the AW polynomials are determined by the
system of Bethe equations \cite{Wieg95}
\beqa
\frac{(y_k-a)(y_k-b)(y_k-c)(y_k-d)}{(ay_k-1)(by_k-1)(cy_k-1)(dy_k-1)}=\prod_{l=1,l\neq
k}^{n}\frac{(qy_k-y_l)(qy_ky_l-1)}{(y_k-qy_l)(y_ky_l-q)}\ \qquad
\mbox{for} \qquad k=1,...,n\ .\label{BetheAW}\eeqa
As shown in \cite{Ter03} (see also \cite{Zhed92} for finite
dimensional representations), the Leonard pair ${\textsf
A}^{(1)},{{\textsf A}^*}^{(1)}$ can be written in this basis.
Introducing the so-called Askey-Wilson second order $q-$difference
operator
\beqa {\mathbb{D}}=\xi(y)(\tau - I) + \xi(y^{-1})(\tau^{-1} - I) +
(1+abcdq^{-1})I\ \label{AWop}\eeqa
where $\tau(y)=qy$ and
\beqa \xi(y)= \frac{(1-ay)(1-by)(1-cy)(1-dy)}{(1-y^2)(1-qy^2)}\
,\nonumber\eeqa
one can define the representation $\pi'$ such that \cite{Ter03}
\beqa \pi'[{\textsf A}^{(1)}]\sim(y+y^{-1})\textsc{P}_{n}(y)
\qquad \mbox{and} \qquad \pi'[{{\textsf
A}^*}^{(1)}]\sim{\mathbb{D}}\textsc{P}_{n}(y)\ . \eeqa
Using the explicit form (\ref{AWop}) in (\ref{spectralN}), setting
$abcd=q$, $c_0=1$ and rescaling the parameters $\kappa,\kappa^*$
one obtains a second order $q-$difference equation of the form
\beqa {\tilde\kappa}^*\xi(y)\textsc{P}_{n}(qy) +
{\tilde\kappa}^*\xi(y^{-1})\textsc{P}_{n}(q^{-1}y)-\big({\tilde\kappa}^*(\xi(y)+\xi(y^{-1})-{\tilde\kappa}(y+y^{-1})-2)\big)\textsc{P}_{n}(y)=
\Lambda_1^{(1)}\textsc{P}_{n}(y)\ .\eeqa
Similarly to the case discussed above, the eigenvalues
$\Lambda_1^{(1)}$ can be written in terms of the zeros
$y_l,y_l^{-1}$ of the AW polynomials determined by
(\ref{BetheAW}).\vspace{1mm}

As both representations are available, let us consider the first
(finite dimensional representation) (\ref{poly}). Then, it is
worth important to notice that the Bethe equations (\ref{Bethe})
as well as the eigenvalue $\Lambda_1^{(1)}$ are invariant under
the symmetry transformation
\beqa q\leftrightarrow q^{-1}\ ,\qquad v\leftrightarrow v^{-1}\ ,
\qquad \kappa\leftrightarrow \kappa^* \ .\label{duality}\eeqa
Although we will not discuss this remarkable symmetry for general
values of $N$, we expect it will be preserved for higher values of
$N$ too.

\section{Examples}
For several quantum completely integrable models, it is not
difficult to construct {\it non-local} integrals of motion in
terms of ${\cal I}^{(N)}_{2n-1} $. Below, we list some examples
(see also \cite{qDG}).\vspace{1mm}

$\bullet$ {\bf Integrable (XXZ) spin chain models}: In the context
of Sklyanin formalism, integrable boundary conditions for a system
of finite size are encoded in the representations of the
$K-$matrix $K^{(N)}(u)$. From previous analysis, the problem is
reduced to classify all possible (tensor product) representations
of ${\textsf A}^{(N)}$, ${{\textsf A}^*}^{(N)}$ associated with
the tridiagonal algebra (\ref{Talg}). For $q$ not a root of unity,
the representations (\ref{TDinit}) and (\ref{TDN}) are proposed.

All known examples of integrable boundary conditions can be indeed
seen as special cases of (\ref{TDN}). For instance, let us denote
$\pi^{(j)}$ the spin$-j$ irreps. of $U_{q^{1/2}}(\widehat{sl_2})$
with (\ref{homo}). For $N=1$ and the trivial representation we can
define $\pi^{(j=0)}[{\textsf A}^{(1)}]=\epsilon_-$,
$\pi^{(j=0)}[{{\textsf A}^*}^{(1)}]=\epsilon_+$. Then $(id\times
\pi^{(j=0)})[K^{(1)}(u)]$ coincides with the non-diagonal solution
of the reflection equation for the open XXZ spin chain with $M$
sites and non-diagonal boundary conditions found in \cite{DV94}.
The method of deriving the ``boundary quantum group'' generators
\cite{Nepo98} proposed in \cite{DelMac03} was recently applied to
this model \cite{Doik04}. It is then easy to see that the
so-called ``boundary quantum group''  is actually the tridiagonal
algebra (\ref{Talg}), the boundary generators being nothing but a
TD pair of the form ${\tilde{\textsf A}}$ $\!\!\!^{(M)}$,
${\tilde{\textsf A} }^*$ $\!\!\!^{(M)}$ in the spin$-\frac{1}{2}$
representation. According to Section 2, this model possesses $M$
mutually conserved quantities given by (\ref{IN}) for
$\kappa=\kappa^*=1$. Interestingly, the corresponding spectral
problem gives a system of $M$-partial $q-$difference equations.
For $N=2$, the $K-$matrix coincides with the one derived in
\cite{qDG}. One can then construct a XXZ spin chain coupled with a
quantum mechanical system \cite{BasDoi04} associated with
$K^{(2)}(u)$, in which case the dynamical boundary conditions are
associated with the AW relations. For general values of $N$, and
any number of sites $M$, it becomes clear that one obtains a XXZ
open spin chain coupled with $N-$type dynamical boundary
conditions taking $K^{(M+N)}(u)$ in (\ref{tN}). Obviously, the
examples above do not exhaust all possibilities. Indeed, starting
from a different representation for the monodromy matrix
satisfying (\ref{RLL}) one gets new quantum integrable models
which share the same underlying symmetry algebra (\ref{Talg}).
Consequently, from a general point of view this family of systems
with $N$ degrees of freedom possesses
\\
- {\it Local} integrals of motion derived from the expansion
$\ln\big(t(\exp(\lambda)\big)=\sum_{n}{\cal H}_n\lambda^n$;\\
- {\it Non-local} integrals of motion derived from the expansion
(\ref{expandt}).\\
Applying the first relation in (\ref{tN}), it follows that they
are mutually commuting. The existence of such integrable structure
of the form (\ref{IN}), besides the known {\it local} integrals of
motion, allows us to say that these models are
superintegrable.\vspace{1mm}

$\bullet$ {\bf Integrable field theory}: Although the construction
of Section 3 can not be directly applied to the case $N\rightarrow
\infty$, it is however possible to identify the conserved
quantities (\ref{IN}) in known quantum integrable models. For
instance, as shown in \cite{Ber91} the sine-Gordon model has a
$U_{q_0}(\widehat{sl_2})$ symmetry generated by non-local
conserved charges (up to an overall scalar factor)
${Q_\pm},{{\overline Q}_\pm}$ satisfying the defining relations
(\ref{Uqsl2}). In this model, the particle spectrum consists of
soliton/anti-soliton and breathers. Usually, each particle is
associated with an asymptotic scattering state $|\theta;m\rangle$
with rapidity $\theta$ and topological charge $2m$ where $-j\leq m
\leq j$. In this representation, the charges act as (\ref{homo})
setting $v=\exp((2/\betah^2-1)\theta)$ with coupling constant
$\betah^2$. On a general asymtotic $p-$particle states, they act
as $\Delta^{(p)}(Q_\pm)$, $\Delta^{(p)}({\overline Q}_\pm)$
\cite{Ber91}. Following \cite{qDG} and the results of Section 2,
the sine-Gordon model or its boundary version (for $c_0=1$)
\cite{GZ94} possess $N\rightarrow \infty$ non-local conserved
charges of form (\ref{IN}) in involution. The corresponding
spectral problem is, in general, rather complicated. However,
acting on a $p-$particle state the spectral problem associated
with (\ref{IN}) truncates to the one associated with the conserved
charges ${\cal I}^{(\infty)}_{2n-1}$, $n=1,...,p$. In particular,
for a single $1-$particle state the corresponding eigenstates are
expressed in terms of Askey-Wilson $q$-orthogonal
polynomials.\vspace{1mm}

$\bullet$ {\bf The limit $q\rightarrow 1$ and the Onsager
algebra}: For $q=1$ and $c_0=1$, the structure (\ref{Talg})
becomes isomorphic to the Dolan-Grady one \cite{DG}, with a
trivial coproduct. In the homogeneous gradation of (\ref{TDinit}),
it is easy to show that the loop algebra $\tilde{sl_2}$ occurs
explicitly as ${\textsf A}\rightarrow E_++E_-$ and ${\textsf
A}^*\rightarrow tE_++t^{-1}E_-$, $t\in \mathbb{C}$, with
$[E_+,E_-]=2H$, $[H,E_\pm]=\pm E_\pm$. In terms of the Onsager
algebra \cite{Ons44} generators $A_m,G_n$, $n\in \mathbb{Z}$ one
has \cite{Roan91} $A_m=2t^mS_++2t^{-m}S_-$ which gives the
identification ${\textsf A}\rightarrow A_0/2$ and ${\textsf
A}^*\rightarrow A_1/2$. It follows that the AW relations
(\ref{AW}) in this limit become
\beqa \frac{1}{8}[A_0,[A_0,A_1]]=A_1-A_{-1} \qquad \mbox{and}
\qquad \frac{1}{8}[A_1,[A_1,A_0]]=A_{0}-A_2\ . \eeqa
Higher order relations are similarly generated. Also, in this
limit it is easy to show that the mutually commuting quantities
(\ref{IN}) coincide with the well-known ones \cite{Davies}. As we
mentioned in the previous Section, for $N=1$ the eigenfunctions
for (\ref{spectralN}) are related with the Askey-Wilson
$q-$orthogonal polynomials (\ref{polyAW}). In the limit $q=1$,
these eigenfunctions reduce to their ``classical'' analogue, i.e.
Wilson, Jacobi, Racah,... polynomials. In this context, it is
probably not surprising that the Jacobi polynomials arise
\cite{Gehl02} in the study of the spectrum of the chiral Potts
model \cite{Baxter88} at the superintegrable point \cite{Potts}.
Finally, let us point out that the integrable models discussed
above (XXZ, sine-Gordon) at this special point $q=1$ enjoy the
Onsager symmetry algebra.

\section{Concluding remarks}
In this paper, we have proposed a generalization of the
Dolan-Grady construction \cite{DG} based on the tridiagonal
algebraic relations (\ref{Talg}). This construction was possible
due to the realization of the TD pair in terms of tensor products
of $U_{q^{1/2}}(\widehat{sl_2})$, which plays a crucial role in
the analysis. It was argued that the corresponding finite set of
mutually commuting quantities can be either derived by imposing
the algebraic structure (\ref{Talg}), or generated from solutions
(\ref{Kdressed}) of the reflection equation. Finally, the
tridiagonal algebraic symmetry has been exhibited in various
examples of quantum integrable models with finite $N$ (or
infinite) degrees of freedom, which are known to possess already
$N$ {\it local} conserved charges. Having identified the
fundamental generators ${{\textsf A}}^{(N)},{{\textsf A}^*}^{(N)}$
and the corresponding $N$ (or infinite) conserved quantities to
{\it non-local} charges (\ref{IN}) in involution, we conclude that
these models are superintegrable.\vspace{1mm}

The next step is the explicit construction of the corresponding
$q-$Onsager symmetry algebra in the spirit of  \cite{Perk,Davies}
that we leave for further investigation. Let us mention that
beyond the explicit construction of the mutually commuting
quantities ${\cal I}_{2n-1}^{(N)}$ with $1\leq n\leq N$, a new
(finite) set of algebraic relations generalizing the Askey-Wilson
ones (\ref{AW}) can be derived directly from (\ref{RE}). Given
$N$, these relations are responsible of the truncation of the
integrable hierarchy i.e. any quantity ${\cal I}_{2n-1}^{(N)}$ for
$n\geq N+1$ can be expressed in terms of ${\cal I}_{2n-1}^{(N)}$
with $n\leq N$. For $N=2$, it is not difficult to obtain such
algebraic relations satisfied by ${\textsf A}^{(2)},{{\textsf
A}^*}^{(2)}$ using the explicit form of the off-diagonal entries
of $K^{(2)}(u)$. We intend to discuss these relations, as well as
the related spectral problem which arises in the classification of
multivariable $q-$orthogonal polynomials separately. \vspace{1mm}

To conclude, as the Dolan-Grady relations arise explicitly in
various problems \cite{Potts,Ahn,Plyu02} we believe our
construction will provide a new tool in order to derive exact
results in massive quantum integrable models. In this direction,
it would be interesting to find the extension of the
Bazhanov-Lukyanov-Zamolodchikov program \cite{BLZ}.\vspace{0.5cm}

\noindent{\bf Acknowledgements:}  I thank  P. Forgacs, N. Kitanine
for discussions and P. Terwilliger for detailed explanations about
his work. I am also grateful to T. Ito and S. Roan for sending
their preprints, and J.H.H. Perk for communications. Part of this
work is supported by the TMR Network EUCLID ``Integrable models
and applications: from strings to condensed matter'', contract
number HPRN-CT-2002-00325.\vspace{1cm}

\end{document}